\documentclass[runningheads]{llncs}
\usepackage{multirow}
\usepackage{siunitx}
\usepackage{graphicx}
\begin{document}
\title{Automated Computer Evaluation of Acute Ischemic Stroke and Large Vessel Occlusion\thanks{Thanks Hong Kong Hospital Authority for Data Preparation}}
\titlerunning{Automated Computer Evaluation of Acute Ischemic Stroke}
%
\author{Jia You\inst{1} \and
Philip L.H. Yu\inst{1} \and
Anderson C.O. Tsang\inst{2}\and
Eva L.H. Tsui\inst{3}\and
Pauline P.S. Woo\inst{3}\and
Gilberto K.K. Leung\inst{2}
}
\authorrunning{J. YOU et al.}
%
\institute{Department of Statistics and Actuarial Science, The University of Hong Kong, Hong Kong \and
Division of Neurosurgery, Department of Surgery, The University of Hong Kong, Hong Kong  \and
Department of Statistics and Workforce Planning, Hospital Authority, Hong Kong\\
\email{\{plhyu\}@hku.hk}}
\maketitle              
\begin{abstract}
Large vessel occlusion (LVO) plays an important role in the diagnosis of acute ischemic stroke. Identifying LVO in the early stage of admission would significantly lower patients’ probability of suffering from severe effects of stroke or even save their lives. In this paper, we utilized both structural and imaging data from all recorded acute ischemic stroke patients in Hong Kong. Total 300 patients (200 training and 100 testing) are used in this study. We established three hierarchical models based on patients’ demographic data, clinical data and features obtained from computerized tomography (CT) scans. The first two stages of modeling are merely based on demographic and clinical data. Besides, the third model utilized extra CT imaging features obtained from deep learning model. The optimal cutoff is determined at the maximal Youden index based on 10-fold cross-validation. With both clinical and imaging features, the Level-3 model achieved the best performance on testing data. The sensitivity, specificity, Youden index, accuracy and area under the curve (AUC) are 0.930, 0.684, 0.614, 0.790 and 0.850 respectively.

\end{abstract}

\section{Introduction}
Acute ischemic stroke (AIS) has becoming a leading cause of morbidity and mortality worldwide, and recent advances in endovascular thrombectomy (EVT) for treatment of AIS caused by large vessel occlusion (LVO) have been widely accepted around the world \cite{b1}.  Similar to intravenous thrombolysis, rapid access to EVT remains paramount that the effectiveness of EVT can be largely ensured if patients initiated within 6 hours from symptom onset than late diagnosed \cite{b1}. Prehospital care focuses on rapid identification of life-threatening emergencies and primary transport to a hospital ideally suited to care for that patient, thereby avoiding the lengthy time delays of interfacility transfers \cite{b2}. 

Recent decades have witnessed the development of prehospital LVO prediction scales in order to differentiate LVO from milder strokes and thus allow paramedics to make rapidly diagnosis in the prehospital setting. Popular scales include the 3-item Stroke Scale \cite{b3}, the Los Angeles Motor Scale \cite{b4}, the Rapid Arterial Occlusion Evaluation Scale \cite{b5}, the Cincinnati Prehospital Stroke Severity Scale \cite{b6}, the Field Assessment Stroke Triage for Emergency Destination \cite{b7} and the Prehospital Acute Stroke Severity \cite{b8}. Some of the above scales specifically aim to identify stroke patients with LVO rather than all AIS patients. These scales are converted from National Institutes of Health Stroke Scale (NIHSS) items, a criterion standard for stroke. The algorithm of these scales is mainly based on the hypothesis of linear correlation between patients’ clinical features and the presence of stroke \cite{b3}. The drawback of these measurements is from the ignorance of some potential stroke-related features, such as age and clinical history factors.

Compared with previous standard scales, this study will combine multiple demographic data, clinical structure data and CT imaging data, and construct a model for LVO prediction. The model will use eXtreme Gradient Boosting (XGBoost) classifier, which is an efficient and scalable implementation of gradient boosting framework by J. Friedman \cite{b9}, \cite{b10}. XGBoost is now a widely used and popular machine learning technique among data scientists’ communities. It is an ensemble technique that builds the model in a stage-wise method that new models are added to correct the errors made by the previously trained models. In the past several years, the deep learning has been widely applied to computer vision tasks, demonstrating the state-of-the-art performances. There are variety of applications in different medical imaging problems as well. In this study, we also adopted deep learning model as a feature extractor to brain CT scans.

\vspace{-.3cm}
\section{Methods}
\vspace{-.2cm}

\subsection{Study Population \& Data Preparation}
\vspace{-.1cm}
The Hong Kong Hospital Authority’s Clinical Management System (CMS) has well-established records of all patients admitted to the public hospitals for all types of acute ischemic stroke. Patients selected in this study need to satisfy all of the following inclusion criteria: (a) aged 18 years or above; (b) with a principal diagnosis of ‘cerebral embolism with mention of cerebral infarction’ (ICD9cm= 434.11) or ‘cerebral artery occlusion, unspecified with mention of cerebral infarction’ (ICD9cm= 434.91); (c) emergency admission; and (d) with a computer tomography (CT) brain scan performed within 24 hours of AED admission. For other detailed criteria and sampling properties please refer to the paper \cite{b11}.

Total 300 subjects were selected and then were randomly split into 200 for model training and 100 for model testing. The data in this study contains basic demographic data, structural clinical data, discharge notes and corresponding CT scans. The demographic data contain some basic information of patients, such as gender and age. The structural clinical data include patients’ records of the pre-existing clinical records, other clinical symptoms and signs at A\&E admission, such as the existence of diabetes mellitus, hypertension and smoking history. Discharge notes are reports written by doctors at the time of discharge from the A\&E department or hospital. All patients have their corresponding brain CT scans as well.

The diagnosis of anterior circulation LVO was independently verified by 2 cerebrovascular disease specialists based on available admission notes, neuroimaging including CT and discharge record. Any discrepancies were resolved by consensus. Among the 300 patients, 130 suffered from LVO and rest 170 were not. The existence of hyperdense middle cerebral artery (MCA) sign is direct visualization of thromboembolic material within the lumen, which is a critical impact factor when making judgment at diagnosis for LVO. The hyperdense MCA dot sign was verified by 2 cerebrovascular disease specialists, as well. Then, the segment label was manually drawn through software FSL \cite{b12}. The CT images have similar quality, spatial resolution and field-of-view. The in-plane resolution is 0.426*0.426 mm. The slice thickness is 5.0 mm for most cases. Each axial slice has identical size of 512*512 pixels. There are 74 cases have witnessed hyperdense MCA dot signs.

\vspace{-.3cm}
\subsection{Hierarchical Modeling}

The present study aimed to develop machine learning models for LVO prediction based on a data hierarchy of three different levels.  Level-1 model only utilizes some basic features that can be easily obtained by every individual. These features include age, gender, the existence of speech deficits, facial weakness, left-facial weakness, right-facial weakness, limb weakness, left-side weakness and right-side weakness.

In additional to the features used in Level-1, Level-2 includes all structural clinical data, such as the existences of diabetes mellitus and hypertension, whether the patient is or was a smoker, current smoker (or quit $<=$ 2 years), previous smoker (quit $>$ 2 years), diastolic and systolic blood pressure, Glasgow coma scale (GCS), and the corresponding sub-scales of eye, verbal and motor. Some previous diagnoses of atrial fibrillation, atherosclerosis and valvular heart disease were included as well. Many of these features contain missing values. 

Together with the features used in Level-1 and Level-2 models, Level-3 model uses image features from patients’ CT scans. Since the hyperdense MCA dot sign is an important factor in the diagnosis of LVO \cite{b13}, we aimed to localize the MCA in level-3 model. Deep learning model works as a feature extractor to obtain additional useful features from CT scans. Each patient has 26 to 30 CT images, and we prefer to choose the slice contains the largest segmented MCA dot sign, which is considered contains more relevant information for LVO. The selected image will be fed into the well-trained architecture, and total 16384 features will be selected for each patient. All features were divided into two subgroups based on whether the patient suffered from LVO. Then, applying a two-sample t-test to each feature, we could identify significant features that can distinguish between the LVO and non-LVO. The top-10 image features with the smallest p-values were extracted and combined with the Level-1 and Level-2 features to build the Level-3 models for the final LVO prediction.

We chose extreme gradient boosting model as our classifier, which superior to off-the-shelf classifiers such as random forest and support vector machine (SVM) \cite{b14}. Another important factor for the choice of XGBoost classifer is the consideration for missing data. Compared with other traditional machine learning methods, XGBoost is capable to handle missing values automatically. Thus, data imputation is not required, and the model can give a robust prediction when new observations contain incomplete features. 

\vspace{-.3cm}
\subsection{Deep Learning}
\vspace{-.1cm}

We applied deep learning in the Level-3 model since the hyperdense MCA dot signs have large contribution to patients’ LVO diagnosis. The proposed architecture belongs to the category of fully convolutional networks (FCN) \cite{b15} that extends the convolution process across the entire image and predicts the segmentation mask as a whole. The architecture works as a feature extractor that the high-level features at the end of encoding part will be subtracted as shown in Fig. 1. The encoding part resembles a traditional convolutional neural networks (CNN) that extract a hierarchy of image features from low to high complexity. The decoding part then transforms the features and reconstructs the segmentation label map from coarse to fine resolution. The model contains skip connections, which is similar to the U-net architecture \cite{b16}. 

\vspace{-.5cm}
\begin{figure}[htbp]
\centerline{\includegraphics[width=11cm,height=7cm]{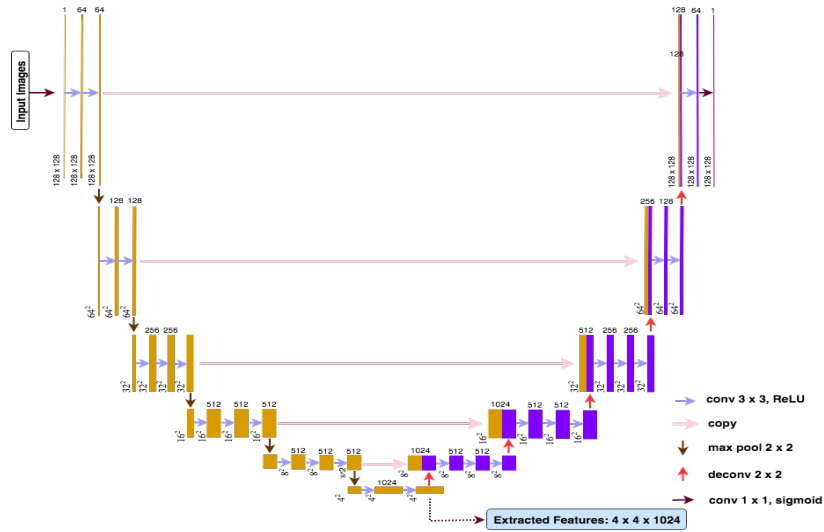}}
\vspace{-.2cm}
\caption{Unet Architecture}
\label{fig}
\end{figure}
\vspace{-.5cm}

The CT scans were preprocessed using the fully automatic pre-processing pipeline through FSL and Nibabel library. As shown in preprocessing flow chart (Fig. 2), the first step is brain extraction to strip the skulls. In the second step, all CT scans are rotated and translated through a rigid-body 2D registration procedure in order to make sure all brains within images are horizontally symmetric. All the MCA dot signs have H.U. index between 30 and 70; thus in step 3, a threshold of 20 to 80 is utilized in order to eliminate the irrelevant image information. Since the hyperdense MCA dot signs largely course in a plane perpendicular to the transverse plane of imaging, the recognition of the MCA dot signs can be localized within a specified area of the scans. To better specify the region where MCA dot sign, we localize a bounding box in step 4. The colored bounding box has size of 128*128; while two colors indicating left and right hemispheres. Besides, given clinical information for different side of limb weakness, we can then extract the infracted hemisphere as shown in the last step. 

\vspace{-.5cm}
\begin{figure}[htbp]
\centerline{\includegraphics[width=8.5cm,height=10cm,keepaspectratio]{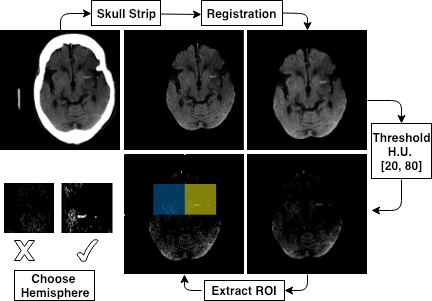}}
\vspace{-.2cm}
\caption{Preprocessing Flow Chart}
\label{fig}
\end{figure}
\vspace{-1cm}

\subsection{Performance Evaluation}
To evaluate the performances of these models, 10-fold cross validation method was adopted. The prediction performance was evaluated by sensitivity, specificity, Youden Index \cite{b17} and area under the curve (AUC) of receiver operating characteristics (ROC) for LVO classification in all three levels.
Aiming to minimize both the sensitivity and specificity of the fitted predictive model, the cutoff was chosen on Youden index, $\gamma$, which is derived from sensitivity and specificity and denotes a linear correspondence balanced accuracy, given as:
\begin{equation}
\gamma = sensitivity + specificity - 1 \label{eq}
\end{equation}
Youden index has been commonly used to evaluate predictive model performance and has shown good performance on model assessment especially for imbalanced dataset \cite{b18}. The best cutoff was obtained through the indication of largest Youden index based on a 10-fold cross validation.

\vspace{-.2cm}

\section{Results}
\vspace{-.3cm}

Besides imaging features in Level-3, the first two models involved total 24 features. To have an intuitive view of all variables, student t-test and Pearson’s Chi-square test were done for continuous and discrete variables, respectively. Based on these tests, a wide range of factors are found to be associated with an increased risk of LVO. These include demographic factors (older age and female), clinical symptoms and signs at A\&E admission (current or previous smokers, existence of left or right facial weakness and limb weakness), clinical structural data and testing scores (systolic blood pressure, GCS, corresponding Verbal, Motor and Eye test scores) and historical clinical records (previous diagnosed cardioembolism, atrial fibrillation and atherosclerosis).

Compared with non-LVO stroke, LVO patients are more likely female than male (67.4\% vs 31.6\%, p=0.003) and older age (mean of 80.5 years vs 71.4 years, p$<$0.001). Glasgow coma scale is also an important indicator for LVO that the lower indcates higher the chance to be LVO of anterior circulation (mean of 10.7 vs 13.68, p$<$0.001); besides, its corresponding eye, verbal and motor testing scores all play critical roles, having small p-values. The prevalence rates in LVO patients are higher than non-LVO for those patients suffering limb weakness (99.2\% vs 74.1\%, p$<$0.001), facial weakness (31.5\% vs 25.9\%, p$<$0.001) and previously atrial fibrillation (36.9\% vs 18.8\%, p$<$0.001). There is no difference in the prevalence of speech deficits, diabetes mellitus, hypertension, prior stroke and ischemic heart disease between the LVO and non-LVO groups. Due to the limitation of paper space, please refer another published paper \cite{b11} for detailed tables and summary statistics.

\vspace{-.7cm}
\begin{figure}[htbp]
\centerline{\includegraphics[width=5.5cm,height=5.5cm,keepaspectratio]{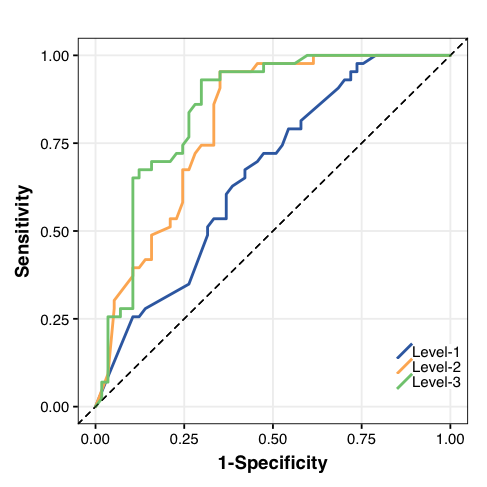}}
\vspace{-.3cm}
\caption{ROC Curve}
\label{fig}
\end{figure}
\vspace{-.6cm}

The testing performance for ROC curve is shown in \textit{Fig. 3} and the detailed measurements are in \textit{Table 1}. Based on the results, the Level-3 model achieves the best performance with all demographical, clinical and imaging features involved in training. Coordinating with the ROC curve in \textit{Fig. 3}, \textit{Table 1} witnesses a great improvement from Level-1 model to Level-2 model regarding all the evaluation measurements. This indicates the clinical symptoms and testing scores at A\&E made great contribution in LVO prediction. Involved with extra imaging features, Level-3 model does not significant improve based on Level-2 model. This is mainly due to the extra features within Level-3 models do not provide enough contributions. According to the Level-2 results, the sensitivity is 93.0\%, while the specificity is 64.9\%, indicating most failed predictions subjects are false positive cases. We aimed to reduce the specificity by adding extra information of MCA dot signs; however, by checking all those false positive subjects are not diagnosed having MCA dot signs. Hence, the features extracted through deep learning model cannot have extraordinary contribution to amend those wrongly predicted cases.
\vspace{-.5cm}
\begin{table}[htbp]
\caption{Testing Performance of all Three-Level Models}
\vspace{-.5cm}
\begin{center}
\centering
\begin{tabular}{|c|c|c|c|c|c|}
\hline
 \scriptsize{\textbf{Hierarchical}} & \scriptsize{\textbf{Sensitivity}} & \scriptsize{\textbf{Specificity}} & \scriptsize{\textbf{Youden}} & \scriptsize{\textbf{Accuracy}} & \scriptsize{\textbf{AUC}}  \\
  \scriptsize{\textbf{Modelling}} & \scriptsize{\textbf{(\%)}} & \scriptsize{\textbf{(\%)}} & \scriptsize{\textbf{Index (\%)}} & \scriptsize{\textbf{(\%)}}&   \\
 \hline
\textbf{Level-1} & {69.8} & {57.9}  & {27.7} & {63.0}  & {0.647}    \\
\hline
\textbf{Level-2} & {93.0} & {64.9}  & {57.9} & {77.0}  & {0.809}    \\
 \hline
 \textbf{Level-3} &  \textbf{93.0} &  \textbf{68.4}  &  \textbf{61.4} &  \textbf{79.0}  &  \textbf{0.850}    \\
 \hline
 \end{tabular}
\label{tab1}
\end{center}
\end{table}
\vspace{-.8cm}

Among the 100 testing subjects, 24 patients were diagnosed to contain MCA dot signs, and the deep learning model is able to correctly predict 22 cases with only 2 subjects missed. Those the sensitivity is fairly good, the specificity is not quite satisfied, which is largely due to some false positive segmented dots.  These inaccurate segmentations are largely due to proximity of bone and the similarity to normal age related vascular calcification.

\vspace{-.3cm}
\section{Discussion}
\vspace{-.3cm}

It is a common problem that patients' demographic and structural clinical data contain missing values. We initially have applied K-nearest neighbors method to impute the missing values, and implemented other machine learning techniques. However, missing value imputation does not make sense in real case application and prediction. The model need to contain all the training data in case to the patient has any incomplete attributes. Regarding this situation, only decision tree and XGBoost methods can handle missing value and take into consideration for the model's complexity, we finally chose XGBoost as the classifier.

It is known that obtain adequate amount of medical data is not easy, especially given the demand for well-diagnosed labels provided by expert clinicians. The deep learning is a data-hungry algorithm that requires huge quantitates of images to keep updating the weights. In this study, among total 300 subjects only 74 subjects contain MCA dot signs, and among these cases, merely 89 slices of scans enclose ground truth. The further study may involve more data to enhance our model’s robustness. 

\vspace{-.3cm}
\section{Conclusion}
\vspace{-.3cm}

In this paper, we present three hierarchical machine learning models that capable to fully automatically predict large vessel occlusion for suspected acute ischemic stroke patients. Combining all features from patients’ demographical, clinical information and brain CT imaging, our final model can give a promising result compare previous methods. Besides, the data involved in this study represents of territory-wide emergency and in-patient healthcare services for a population of 7.3 million. Thus, it can work as an effective tool in assistant to doctors in making quick and accurate diagnosis at the early stage of patients’ admission.


\begin{thebibliography}{8}
\vspace{-.3cm}

\bibitem{b1} Powers, W.J., Derdeyn, C.P., Biller, J., Coffey, C.S., Hoh, B.L., Jauch, E.C., et al.: 2015 American Heart Association/American Stroke Association focused update of the 2013 guidelines for the early management of patients with acute ischemic stroke regarding endovascular treatment. Stroke \textbf{46}(10), 3020-3035 (2015)

\bibitem{b2} Prabhakaran, S., Ward, E., John, S., Lopes, D.K., Chen, M., Temes, R.E., et al.: Transfer delay is a major factor limiting the use of intra-arterial treatment in acute ischemic stroke. Stroke \textbf{42}(6), 1626-1630. (2011)

\bibitem{b3} Singer, O. C., Dvorak, F., du Mesnil de Rochemont, R., Lanfermann, H., Sitzer, M., Neumann-Haefelin, T.: A simple 3-item stroke scale: comparison with the National Institutes of Health Stroke Scale and prediction of middle cerebral artery occlusion. Stroke \textbf{36}(4), 773-776 (2005)

\bibitem{b4} Nazliel, B., Starkman, S., Liebeskind, D.S., Ovbiagele, B., Kim, D., Sanossian, N., et al.: A brief prehospital stroke severity scale identifies ischemic stroke patients harboring persisting large arterial occlusions. Stroke \textbf{39}(8), 2264-2267 (2008)

\bibitem{b5} Pérez de la Ossa, N., Carrera, D., Gorchs, M., Querol, M., Millán, M., Gomis, M., et al.: Design and validation of a prehospital stroke scale to predict large arterial occlusion: the rapid arterial occlusion evaluation scale. Stroke \textbf{45}(1), 87-91 (2014)

\bibitem{b6} Katz, B. S., McMullan, J. T., Sucharew, H., Adeoye, O., Broderick, J. P.: Design and validation of a prehospital scale to predict stroke severity: Cincinnati Prehospital Stroke Severity Scale. Stroke STROKEAHA-115 (2015).

\bibitem{b7} Lima, F.O., Silva, G.S., Furie, K.L., Frankel, M.R., Lev, M.H., Camargo, É.C., et al.: Field assessment stroke triage for emergency destination: a simple and accurate prehospital scale to detect large vessel occlusion strokes. Stroke \textbf{47}(8), 1997-2002 (2016)

\bibitem{b8} Hastrup, S., Damgaard, D., Johnsen, S. P., Andersen, G.: Prehospital acute stroke severity scale to predict large artery occlusion: design and comparison with other scales. Stroke STROKEAHA-115 (2016)

\bibitem{b9} Friedman, J., Hastie, T.,Tibshirani, R.: Additive logistic regression: a statistical view of boosting. The annals of statistics \textbf{28}(2), 337-407 (2000)

\bibitem{b10} Friedman, J. H.: Stochastic gradient boosting. Computational Statistics \& Data Analysis \textbf{38}(4), 367-378 (2002)

\bibitem{b11} Tsang, A.C.O., You, J., Li, L.F., Tsang, F.C.P., Woo, P.P.S., Tsui, E.L.H., et al.: Burden of large vessel occlusion stroke and the service gap of thrombectomy: a population-based study using a territory-wide public hospital system registry. To appear in International Journal of Stroke (2018)

\bibitem{b12} Jenkinson, M., Beckmann, C. F., Behrens, T. E., Woolrich, M. W., Smith, S. M. (2012). Fsl. Neuroimage \textbf{62}(2), 782-790.

\bibitem{b13} Lim, J., Magarik, J. A., Froehler, M. T.: The CT‐Defined Hyperdense Arterial Sign as a Marker for Acute Intracerebral Large Vessel Occlusion. Journal of Neuroimaging \textbf{28}(2), 212-216 (2018)

\bibitem{b14} Chen, T., Guestrin, C.: Xgboost: A scalable tree boosting system. In: Conference on Knowledge Discovery and Data Mining, pp. 785-794 (2016)

\bibitem{b15} Long, J., Shelhamer, E., Darrell, T.: Fully convolutional networks for semantic segmentation. In: IEEE Conference on CVPR. pp. 3431-3440 (2015)

\bibitem{b16} Ronneberger, O., Fischer, P., Brox, T.: U-Net: convolutional networks for biomedical image segmentation. In: MICCAI. pp. 234–241 (2015)

\bibitem{b17} Youden, W. J.: Index for rating diagnostic tests. Cancer \textbf{3}(1), 32-35 (1950)

\bibitem{b18} Bekkar, M., Djemaa, H. K., Alitouche, T. A.: Evaluation measures for models assessment over imbalanced datasets. Journal Of Information Engineering and Applications, \textbf{3}(10) (2013)

\end{thebibliography}
\end{document}